\documentclass[12pt]{article}

\usepackage{latexsym}

\usepackage[cp1251]{inputenc}

\usepackage{amsmath,amssymb}
\usepackage{indentfirst}

\usepackage[dvips]{graphicx}
\usepackage{epstopdf}

\begin{document}


\title{The Shadow of a Rotating Traversable Wormhole}

\author{
    Petya G. Nedkova\thanks{E-mail:pnedkova@phys.uni-sofia.bg}, Vassil Tinchev\thanks{E-mail:tintschev@phys.uni-sofia.bg}, Stoytcho S. Yazadjiev\thanks{E-mail: yazad@phys.uni-sofia.bg}\\
{\footnotesize  Department of Theoretical Physics,
                Faculty of Physics, Sofia University,}\\
{\footnotesize  5 James Bourchier Boulevard, Sofia~1164, Bulgaria }\\
}

\date{}

\maketitle

\begin{abstract}
We explore the shadow of certain class of rotating traversable wormholes within classical general relativity. The images depend on the angular momentum of the wormhole, and the inclination angle of the observer. We compare the results with the case of the Kerr black hole. For small angular momenta the shadows for the two solutions are nearly identical, however with the increasing of the angular momentum they start to deviate considerably.
\end{abstract}

\section{Introduction}
Wormholes are one of the most interesting predictions of general relativity. They represent solutions to the field equations which are topologically non-simply connected. Thus, they are interpreted as tunnels in spacetime connecting separated parts of our universe, or regions in two different universes, if a multiverse scenario is adopted. Early examples of wormholes date back to the works of Flamm \cite{Flamm:1916}, Einstein and Rosen \cite{Einstein:1935}, and Wheeler \cite{Wheeler:1955}, and the modern development of the area was triggered in a great extend by the idea of Morris and Thorne that traversable wormhole can be constructed \cite{Morris:1988}. Traversability means physically that a human being would be able to pass intact through the tunnel of the wormhole in both directions and in a reasonable time. Consequently, such a wormhole should contain no spacetime singularities or horizons, and induce bearable tidal forces. If existing or possible to construct, it would enable fascinating applications like traveling between distant galaxies, or even time travel.

The typical approach in obtaining wormhole solutions is constructing a singularity-free metric which describes the appropriate wormhole geometry, and investigating afterwards what kind of matter should be present in order for the constructed metric to satisfy the field equations. Following this line of reasoning Morris and Thorne explored the static traversable wormholes with no time dependence, and  concluded that they cannot be build only by ordinary matter \cite{Morris:1988}. All known classical forms of matter possess a stress-energy tensor which satisfies certain energy conditions \cite{Hawking:1973}. However, to be viable solutions to the Einstein equations, Morris-Thorne wormholes should contain matter with stress-energy tensor violating the null energy condition, and consequently all the other ones. Rotating axially symmetric solutions describing traversable wormholes were also obtained subsequently \cite{Teo:1998}, including such with time-dependent angular velocity \cite{Kuhfittig:2003a}. In a similar way it was demonstrated that they require violation of the energy conditions. Therefore, wormholes are often called exotic solutions of general relativity, since they should contain some 'exotic' form of matter.

Various attempts were made to minimize the violation of the energy conditions by constructing wormhole configurations were the amount of exotic matter is arbitrary small, or it is restricted only to particular regions \cite{Visser:1989}. Thus, the eventual traveler could possibly not encounter it when passing through the wormhole. In another line of research it is argued that wormholes should be considered in semi-classical regime. Then, the violation of the energy conditions is not unusual since it occurs also in other quantum systems, like in the Casimir effect, or Hawking evaporation. This motivated the construction of a number of wormhole solutions within the semi-classical gravity \cite{Sushkov:1992}. Attempts were made also to relate the exotic matter supporting wormholes with cosmological models. According to the most popular cosmological scenario, the universe is composed predominantly by some negative pressure substance called dark energy. Certain candidates for dark energy, like the phantom energy, also violate the null energy condition, thus resembling the wormhole case. Inspired by this observation, a series of wormhole solutions containing phantom energy occurred  \cite{Lobo:2005}.

Finally, it should be mentioned that the existence of exotic matter can be  completely avoided by considering alternative theories of general relativity. For example, in dilatonic Gauss-Bonnet gravity, and other higher order curvature theories like $f(R)$ theories wormholes have been constructed without any need of exotic matter \cite{Kanti:2011}.

Besides as an eventual means of interstellar traveling, wormholes arise astrophysical interest as compact objects possibly inhabiting our universe. It is generally considered that the galactic centers contain a supermassive compact object, which is most commonly  believed to constitute a black hole. However, horizonless objects like boson stars, gravastars, and wormholes cannot be currently excluded. Consequently, it is important for future observations to consider tests which can distinguish between wormholes and black holes. Previous developments include the investigation of the gravitational lensing by wormholes \cite{Cramer:1995}, the properties of particle motion in their vicinity \cite{Muller:2008}, and accretion disks in wormhole spacetimes \cite{Harko:2008}. Another feature which can be used to extract physical information by direct observation is the shadow cast by the compact object, or equivalently its apparent shape \cite{Falke:2000}. Experiments suitable for such observations include the Event Horizon Telescope \cite{EH}, which is a system of earth-based telescopes measuring in the (sub)millimeter wavelength, the space-based radio telescopes  RadioAstron and Millimetron \cite{RS}, \cite{Johannsen:2012}, or the space-based X-ray interferometer MAXIM \cite{MX}. In the next few years these missions are expected to reach resolution high enough to observe the shadow of the supermassive compact object at the center of our galaxy or those located at nearby galaxies \cite{Johannsen:2012}.

The existence of a shadow is  characteristic for black hole solutions, and it is thoroughly investigated for the Kerr-Newmann family \cite{Bardeen}-\cite{Hioki:2009}. Shadows of black holes possessing nontrivial NUT-charge were obtained in \cite{Ahmedov:2013}, and black hole solutions within Einstein-Maxwell-dilaton gravity and Chern-Simons modified gravity were considered in \cite{Amarilla:2013},\cite{Amarilla:2010}. The apparent shape of the Sen black hole is studied in \cite{Hioki:2008}. The aim of the current paper is to investigate the apparent shape of a wormhole, and compare the results with the images for the Kerr black hole. Thus, we can draw conclusions on their possible distinction in astrophysical observation. The shadow of a static traversable wormhole within classical general relativity was investigated in \cite{Bambi:2013}. In our work we consider the more general class of rotating traversable wormholes which are described by the general solution found by Teo \cite{Teo:1998}.

The article is organized as follows. In the next section we describe briefly the exact solution representing stationary and axisymmetric traversable wormhole. In section 3 we derive the geodesic equations describing light propagation in its vicinity, and the algebraic equations determining its shadow are obtained in section 4. Finally, we present the images that should be seen by a distant observer for several characteristic angular momenta of the wormhole, and different inclination angles.

\section{Rotating traversable wormhole}

A stationary axisymmetric solution to the Einstein equations describing  rotating traversable wormhole was obtained by Teo \cite{Teo:1998} as a generalization of the static Morris-Thorne wormhole. It is given  by the following metric

\begin{equation}\label{metric}
ds^2=-N^2{\rm d}t^2+\left(1-{b\over r}\right)^{-1}{\rm d}r^2+r^2K^2
\left[{\rm d}\theta^2+\sin^2\theta({\rm d}\varphi-\omega{\rm d}t)^2
\right],
\end{equation}
where $r$, $\theta$ and $\phi$ are spherical coordinates, and the functions $N$, $b$, $K$, and $\omega$ depend only on $r$ and $\theta$. The orbits of the timelike and the spacelike Killing fields are parameterized by the coordinates $t$ and $\phi$.

The function $N$ is frequently called a redshift function since it determines the gravitational redshift. In order for the wormhole to be traversable, it should be finite and nonzero, so that no curvature singularities and event horizons occur.  The function $b$ is the so called shape function and it determines the shape of the wormhole. It is assumed to be non-negative, and contains an apparent singularity at $r=b\geq0$ which corresponds to the throat of the wormhole. The function $b$ is also required to be independent of the coordinate $\theta$ at the throat, i.e $\partial_\theta b(r,\theta)=0$, because otherwise curvature singularity is present. Consequently, for a regular solution the throat represents a 2-dimensional surface located at some constant radius $r=r_0$. Frequently, a further condition is imposed on the function $b$ to ensure that the wormhole possesses the characteristic shape considered by Morris and Thorne (see fig.1 in Ref. \cite{Morris:1988}). It is called the flare-out condition, and arises by studying the embedding of the 2-dimensional cross-section of the solution at constant $t$ and $\theta$ into 3-dimensional Euclidean space. The 2-dimensional surface

\begin{equation}
ds_{(2)}^2=\left(1-{b\over r}\right)^{-1}{\rm d}r^2+r^2K^2
\sin^2\theta d\varphi^2
\end{equation}
'flares out' at $r>r_0$, if the shape function satisfies $\partial_r b(r, \theta)<1$ at the throat \cite{Teo:1998}.

The remaining metric function $K$ is a regular, positive and non-decreasing function determining the proper radial distance $R = rK$, while the function $\omega$ is connected with the angular velocity of the wormhole. To ensure that the metric is nonsingular on the rotation axis $\theta =0$ and $\theta = \pi$ the derivatives of $N$, $K$ and $b$ with respect to $\theta$ should vanish on it.

As a result the described metric represents two identical regions  joined together at the throat $r = b=r_0$. The radial coordinate takes the range $r_0\leq r < \infty$, and the limit $r\rightarrow \infty$ corresponds to the physical infinity . By physical reasons the described wormhole solution is assumed to be asymptotically flat. Therefore, the metric functions should possess the following behavior at $r\rightarrow \infty$

\begin{eqnarray}
&&N = 1 - \frac{M}{r} + {\rm O}\left({1\over r^2}\right), \quad ~~~ K = 1 + {\rm O}\left({1\over r}\right) , \quad \frac{b}{r} = {\rm O}\left({1\over r}\right), \nonumber \\
&&\omega={2J\over r^3}+{\rm O}\left({1\over r^4}\right).
\end{eqnarray}
The constants involved in the asymptotic expansions correspond to the conserved charges of the solution. $M$ determines the mass of the wormhole, while $J$ is equal to its angular momentum.

Except for the described restrictions necessary for the regularity and the physical relevance of the solution, the metric functions $N$, $K$, $b$ and $\omega$ can be chosen at will, and the obtained solution will represent a particular case of rotating traversable wormhole. For our purposes, in the remaining part of article we will consider the class of solutions, when all the metric functions depend only on the radial coordinate $r$. These solutions reduce to the Morris-Thorne wormhole in the limit of zero rotation $\omega = 0$.

\section{Propagation of light in the spacetime of traversable wormhole}

The motion of test particles in a particular spacetime is determined by the corresponding geodesic equations, which follow from the Hamilton-Jacobi equation
\begin{equation}  \label{HJ}
\frac{\partial S}{\partial \lambda}=-\frac{1}{2}g^{\mu\nu}\frac{\partial S}{\partial x^{\mu}}\frac{\partial S}{\partial x^{\nu}}.
\end{equation}

We denote by $\lambda$ an affine parameter along the geodesics, $g_{\mu\nu}$ are the components of the metric tensor, and $S$ is the Jacobi action. In general, the geodesic motion in stationary and axisymmetric spacetime allows two integrals of motion - the energy of the particle $E$  and its angular momentum about the axis of symmetry $L$. If a further conserved quantity is present, the so called Carter constant \cite{Carter:1968}, the Hamilton-Jacobi equation is separable, and it possesses solution of the form

\begin{equation}\label{SHJ}
S=\frac{1}{2}\mu ^2 \lambda - E t + L \varphi + S_{r}(r)+S_{\theta}(\theta),
\end{equation}
where $\mu $ is the mass of the test particle. We denote by $t$ the timelike coordinate, $\varphi$ parameterizes the orbits of the spacelike Killing field, and $S_r(r)$ and $S_\theta(\theta)$ are functions only of the specified coordinates.

If we consider a rotating wormhole solution described by the metric ($\ref{metric}$), in which all the metric functions depend only on the radial coordinate, the Hamilton-Jacobi equation is separable. Using the ansatz ($\ref{SHJ}$) it reduces to the following equations for the functions $S_r(r)$ and $S_\theta(\theta)$

\begin{eqnarray}\label{HJSS}
&&\left(\frac{dS_\theta}{d\theta}\right)^2 = Q - \frac{L^2}{\sin^2\theta}, \nonumber \\
&&\left(1-\frac{b}{r}\right)N^2\left(\frac{dS_r}{dr}\right)^2 = \left(E-\omega L\right)^2 - \left(\mu^2N^2 + Q \frac{N^2}{r^2K^2}\right),
\end{eqnarray}
where $Q$ is the Carter constant. Denoting by $T(\theta)$ and $R(r)$ the expressions in the righthand side of the equations

\begin{eqnarray}\label{RT}
T(\theta) &=& Q - \frac{L^2}{\sin^2\theta}, \nonumber \\
R(r) &=& \left(E-\omega L\right)^2 - \left(\mu^2N^2 + Q \frac{N^2}{r^2K^2}\right),
\end{eqnarray}
the Jacobi action can be obtained in the form
\begin{equation}  \label{SHJ}
S=\frac{1}{2}\mu ^2 \lambda - E t + L \varphi + \int{\sqrt\frac{R(r)}{N^2\left(1-\frac{b}{r}\right)}dr} + \int{\sqrt{T(\theta)}d\theta}.
\end{equation}

The geodesic equations governing a test particle motion are derived from the Jacobi action by setting to zero all its partial derivatives with respect to the constants of motion $\mu$, $E$, $L$ and $Q$. Since we will be interested in photon motion, we should set subsequently the mass of the particle $\mu$ to zero. Thus, we obtain the following equations for the null geodesics in the spacetime of rotating traversable wormhole

\begin{eqnarray}\label{geodesic}
&&\frac{N}{\left(1-\frac{b}{r}\right)^{1/2}}\frac{dr}{d\lambda} = \sqrt{R(r)}, \quad~~~
r^2K^2\frac{d\theta}{d\lambda} = \sqrt{T(\theta)}, \nonumber \\[2mm]
&&N^2\frac{d\varphi}{d\lambda} = \omega(E - \omega L) + \frac{N^2~L}{r^2K^2\sin^2\theta}, \nonumber \\[2mm]
&&N^2\frac{dt}{d\lambda} =E - \omega L.
\end{eqnarray}
The functions $R(r)$ and $T(\theta)$ are given by ($\ref{RT}$) with $\mu =0$, and they should be non-negative for classical motion. The geodesic equations are parameterized by the constants of motions $E$, $L$ and $Q$, but only two of the quantities are independent. We can introduce the ratios

\begin{eqnarray}
\xi = \frac{L}{E},\quad~~~ \eta = \frac{Q}{E^2},
\end{eqnarray}
called impact parameters, and a new affine parameter $\tilde{\lambda}= E\lambda$, and eliminate the energy from the geodesic equations. Thus, the photon motion is parameterized only by $\xi$ and $\eta$. In terms of the impact parameters the functions $R(r)$ and $T(\theta)$ take the form

\begin{eqnarray}
R(r) &=& \left(1-\omega \xi\right)^2 - \eta \frac{N^2}{r^2K^2}, \nonumber \\
T(\theta) &=& \eta - \frac{\xi^2}{\sin^2\theta}.
\end{eqnarray}

\section{The shadow of a wormhole}

We will consider a wormhole connecting two regions of spacetime, such that in one of the regions the wormhole is illuminated by a source of light, and in the other region no sources of light are present in the vicinity of the throat. In the first region photons will propagate most generally on two types of orbits - orbits plunging into the wormhole and passing through its throat, and others scattered away from the wormhole to infinity. A distant observer situated in the first region will be able to see only photons scattered away from the wormhole, and those captured by the wormhole will form a dark spot. This dark region observed on the luminous background is called shadow of the wormhole.

The photon orbits are determined by the impact parameters and for certain values of $\xi$ and $\eta$ a critical orbit exists separating escape and plunge orbits. It corresponds to the boundary of the shadow. We can determine the critical orbit by analyzing the radial geodesic equation, which can be written in the form of an energy-like equation

\begin{eqnarray}
\left(\frac{dr}{d\tilde{\lambda}}\right)^2 + V_{eff} = 1  , \quad V_{eff} = 1 - \frac{1}{N^2}\left(1-\frac{b}{r}\right)R(r),
\end{eqnarray}
by means of an effective potential $V_{eff}$ depending on the impact parameters. The particle will scatter away from the wormhole only if its radial motion possesses a turning point $dr/d{\tilde\lambda} = 0$. Consequently, the critical orbit between escape and plunge motion corresponds to the highest maximum of the effective potential. It is a spherical orbit, meaning that it is located at constant radius, and it is unstable, since a small perturbation in the impact parameters can turn it either to an escape or to a capture orbit. The position of the unstable spherical orbit is determined by the standard conditions for the maximum of the effective potential
 \begin{eqnarray}
 V_{eff} = 1,\quad~~~\frac{V_{eff}}{dr} = 0,\quad~~~\frac{d^2 V_{eff}}{dr^2} \leq 0.
 \end{eqnarray}
Considering the explicit form of the $V_{eff}$ and taking into account that the functions $N$ and $(1-b/r)$ are finite and nonzero outside the throat of the wormhole, the boundary of the shadow can be determined equivalently by the lowest minimum of the function $R(r)$, i.e.

\begin{eqnarray}
R(r) = 0,\quad~~~ \frac{dR}{dr}=0,\quad~~~\frac{d^2R}{dr^2}\geq0.
\end{eqnarray}
Thus, we obtain two algebraic equations for the impact parameters and the radial position of the unstable spherical orbit. They define a relation between the impact parameters $\eta(\xi)$  which should be satisfied on the boundary of the shadow. For convenience we can also represent it in parametric form, expressing  $\xi$ and $\eta$ as a function of the radial position. The following algebraic relations are obtained
\begin{eqnarray}\label{xi}
&&\eta = \frac{r^2K^2}{N^2}(1-\omega\xi)^2, \nonumber\\
&&\xi = \frac{\Sigma}{\Sigma\omega-\omega'}, \quad~~~ \Sigma = \frac{1}{2}\frac{d}{dr}\ln\left(\frac{N^2}{r^2K^2}\right),
\end{eqnarray}
where $(...)'$ denotes differentiation by $r$.  In addition, $\xi$ and $\eta$ should be such that the condition $T(\theta)\geq 0$ to be satisfied, in order to obtain valid classical solutions for the $\theta$-motion geodesic equation.

The derived relations ($\ref{xi}$) define the boundary of the shadow in the impact parameter space. In reality the observer at infinity will see a projection of it at the so called 'observer's sky', i.e. the plane passing through the wormhole and normal to the line connecting it with the observer (the line of sight). The coordinates at this plane, which we will denote by  $\alpha$ and $\beta$, are called celestial coordinates, and they give the apparent position of the image. The celestial coordinates are connected with the geodesic equations as \cite{Bray:1986}

\begin{eqnarray}  \label{alpha}
\alpha&=&\lim_{r\rightarrow \infty}\left( -r^{2}\sin\theta_{0}\frac{d\varphi}{dr}\right), \nonumber \\
\beta&=&\lim_{r\rightarrow \infty}r^{2}\frac{d\theta}{dr},
\end{eqnarray}
where $\theta_0$ is the angle between the rotation axis of the wormhole (the $\varphi$-axis) and the line of sight of the observer, called inclination angle. Considering the geodesic equations ($\ref{geodesic}$) we can deduce explicit expressions for the celestial coordinates for our wormhole solution

\begin{eqnarray}
\alpha &=& -\frac{\xi}{\sin\theta_0}, \nonumber \\
\beta &=& \left(\eta - \frac{\xi^2}{\sin^2\theta_0}\right)^{1/2}.
\end{eqnarray}
If we substitute the impact parameters $\xi$ and $\eta$ with the relations ($\ref{xi}$) determining the boundary of the shadow, we will obtain its apparent image  as seen by an observer at infinity which depends on the angular momentum of the wormhole and the inclination angle of the
observer.

\begin{figure}[htp]
		\setlength{\tabcolsep}{ 0 pt }{\scriptsize\tt
		\begin{tabular}{ cccc }
            \includegraphics[width=4.1 cm]{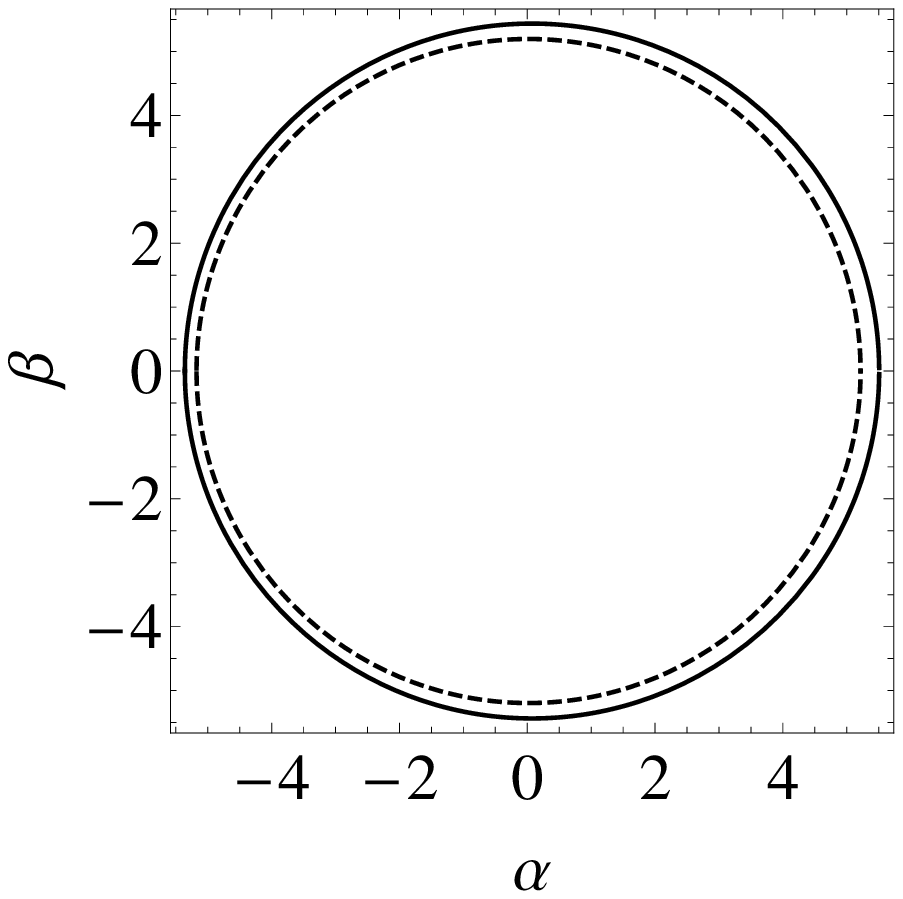} &
            \includegraphics[width=4.1 cm]{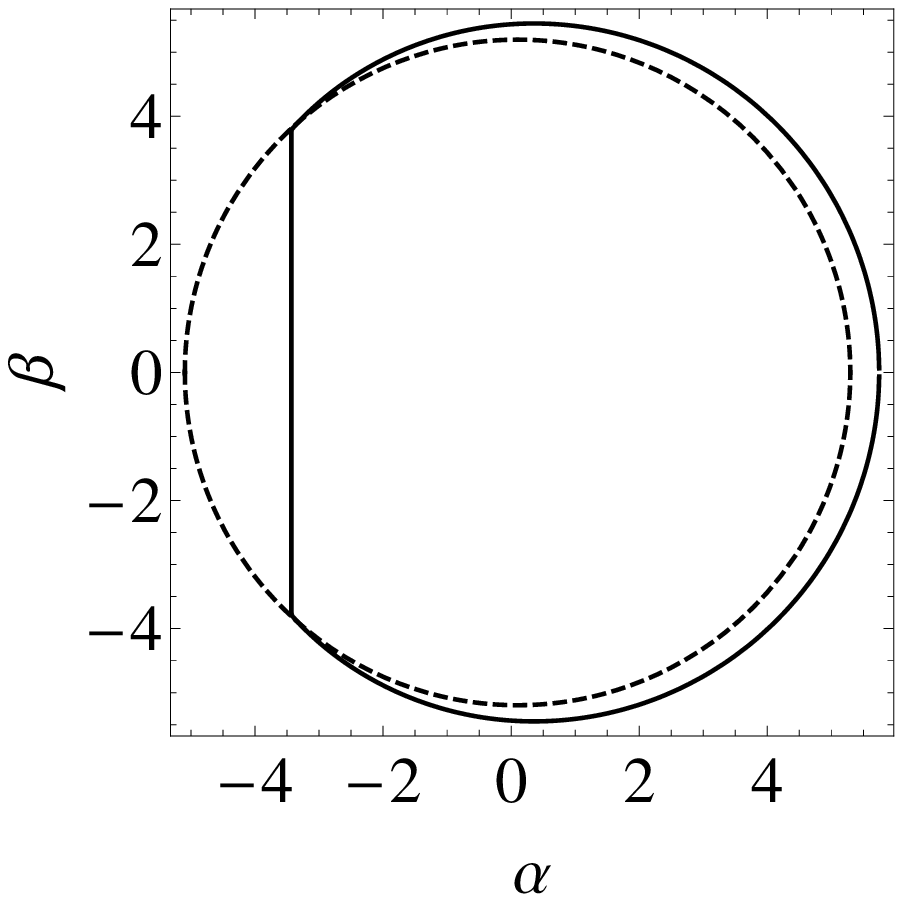} &
            \includegraphics[width=4.1 cm]{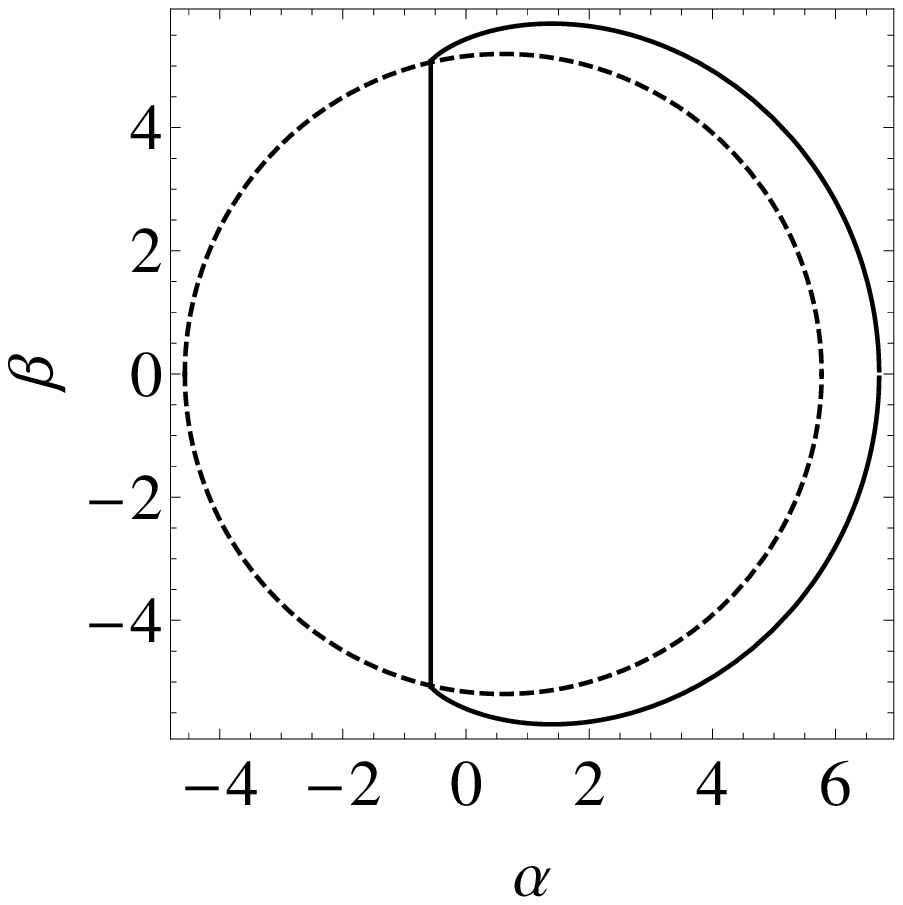} &
			\includegraphics[width=4.1 cm]{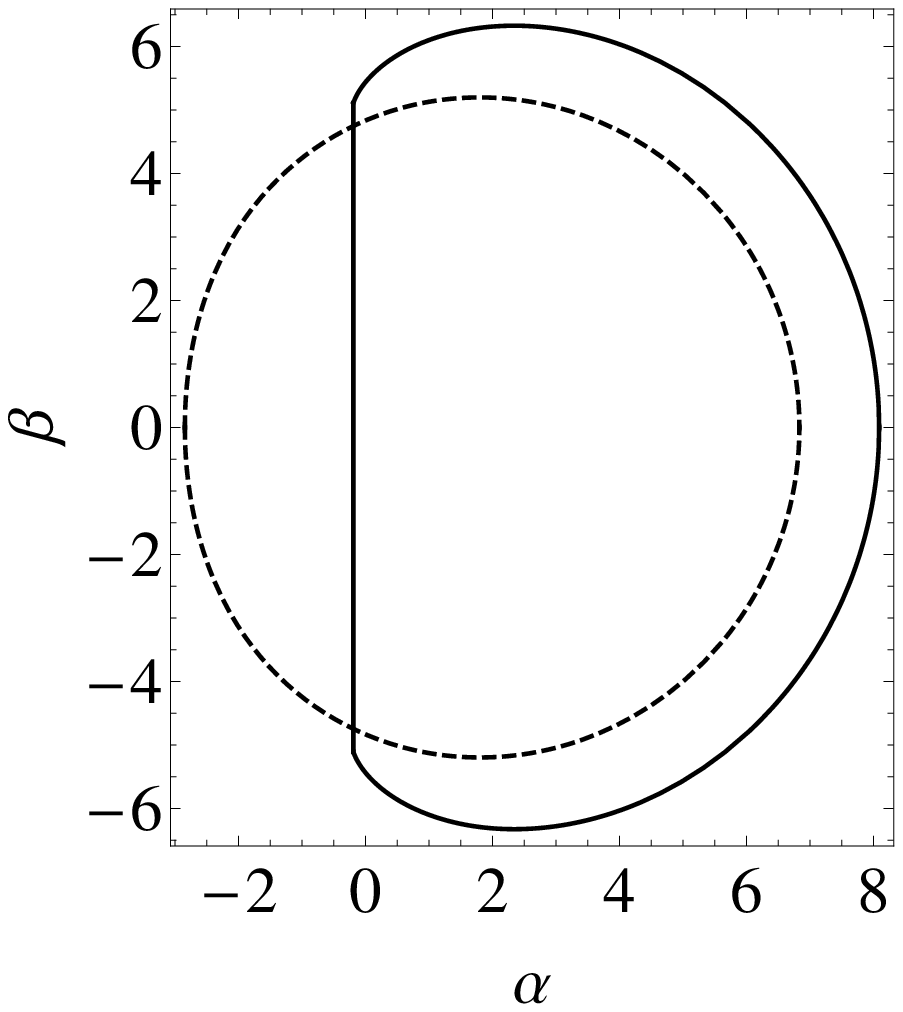} \\
			$J/M^{2}=0.01$, $\theta_{0}=90^\circ$\  &
			$J/M^{2}=0.05$, $\theta_{0}=90^\circ$\  &
            $J/M^{2}=0.3$, $\theta_{0}=90^\circ$\  &
			$J/M^{2}=0.9$, $\theta_{0}=90^\circ$ \\[2mm]
            \includegraphics[width=4.1cm]{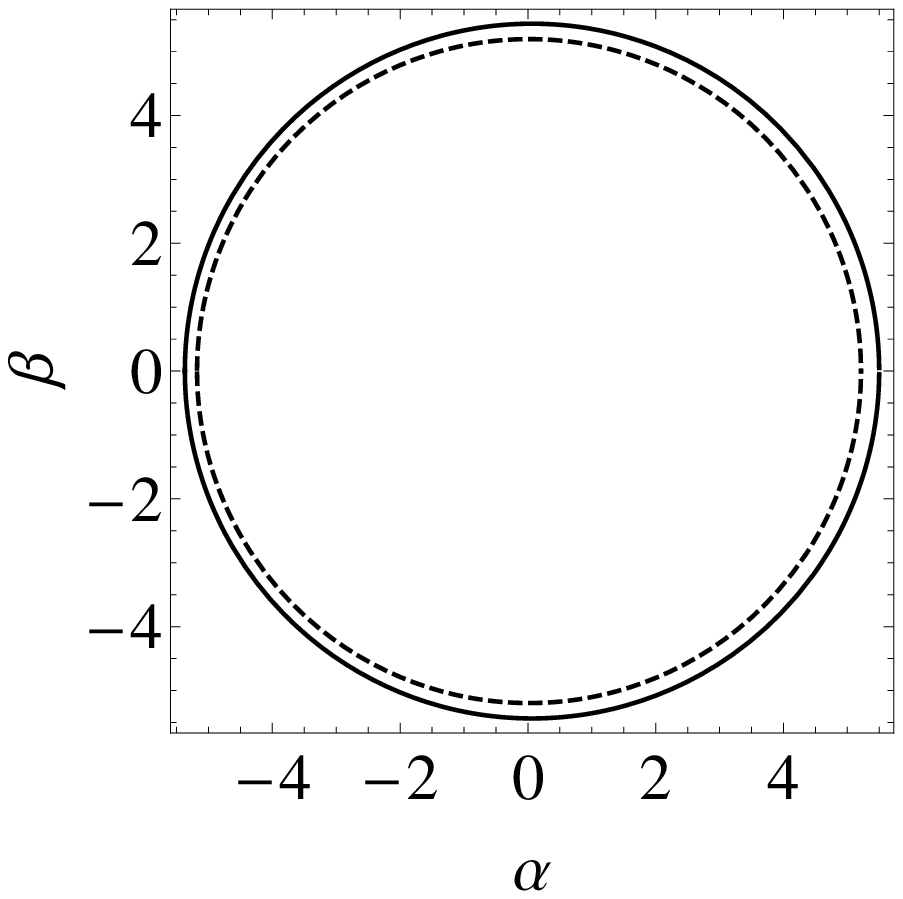} &
            \includegraphics[width=4.1cm]{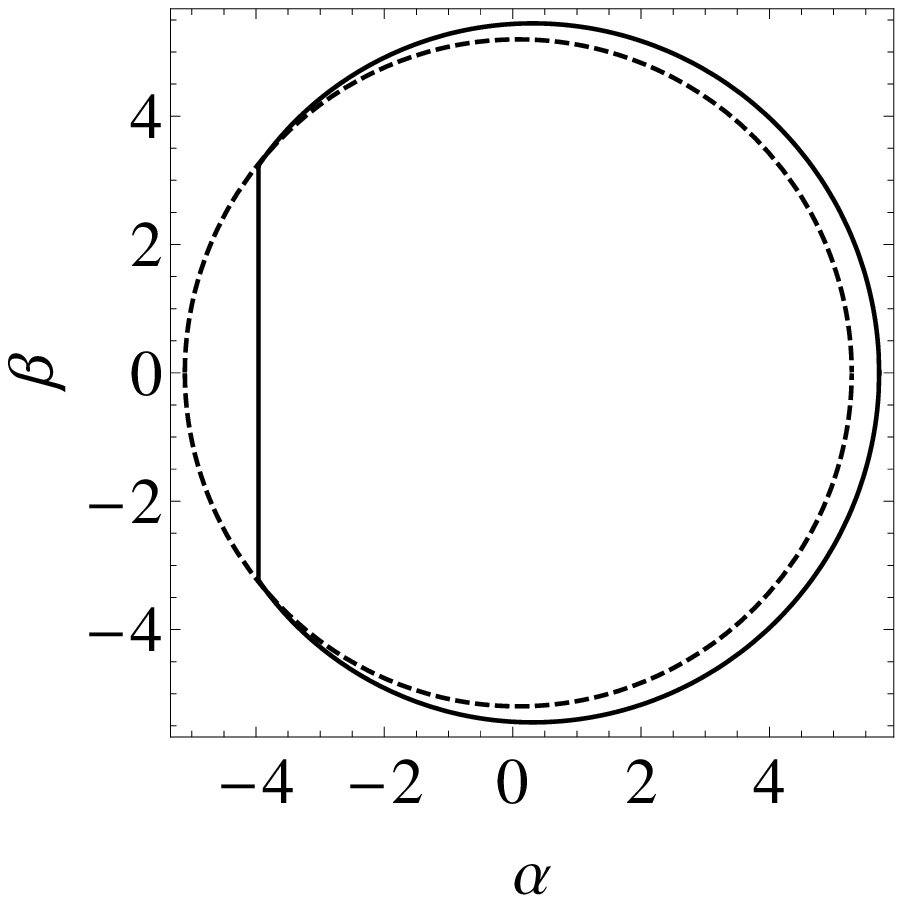} &
            \includegraphics[width=4.1cm]{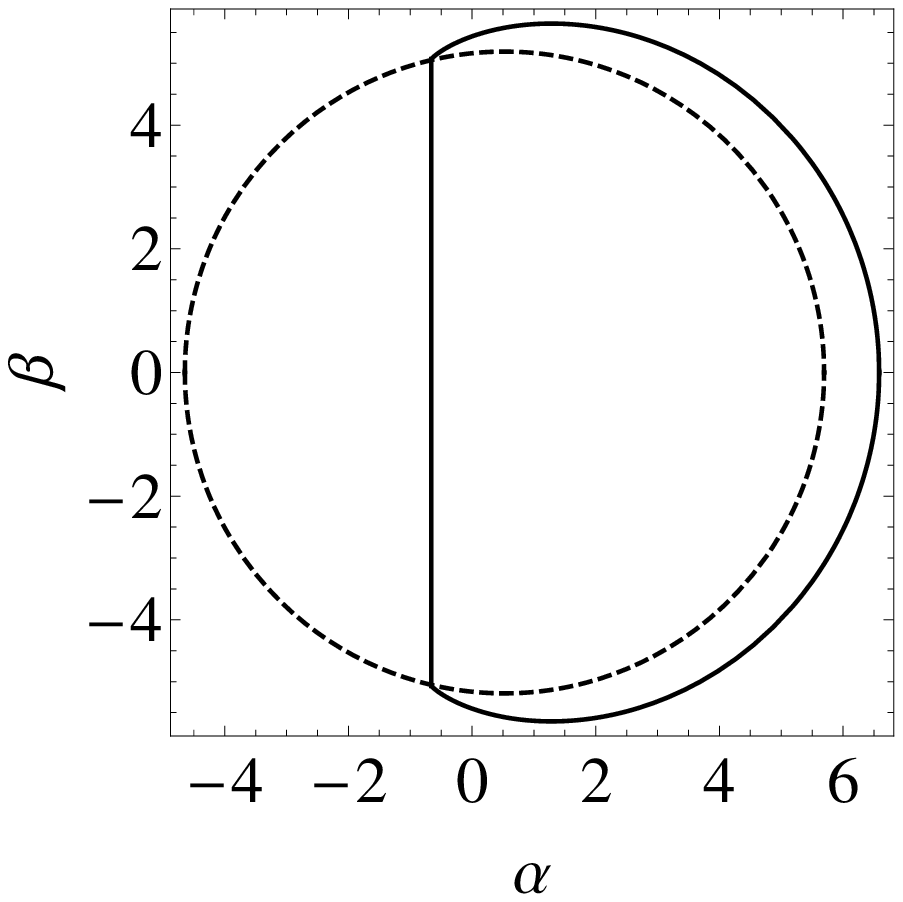} &
			\includegraphics[width=4.1cm]{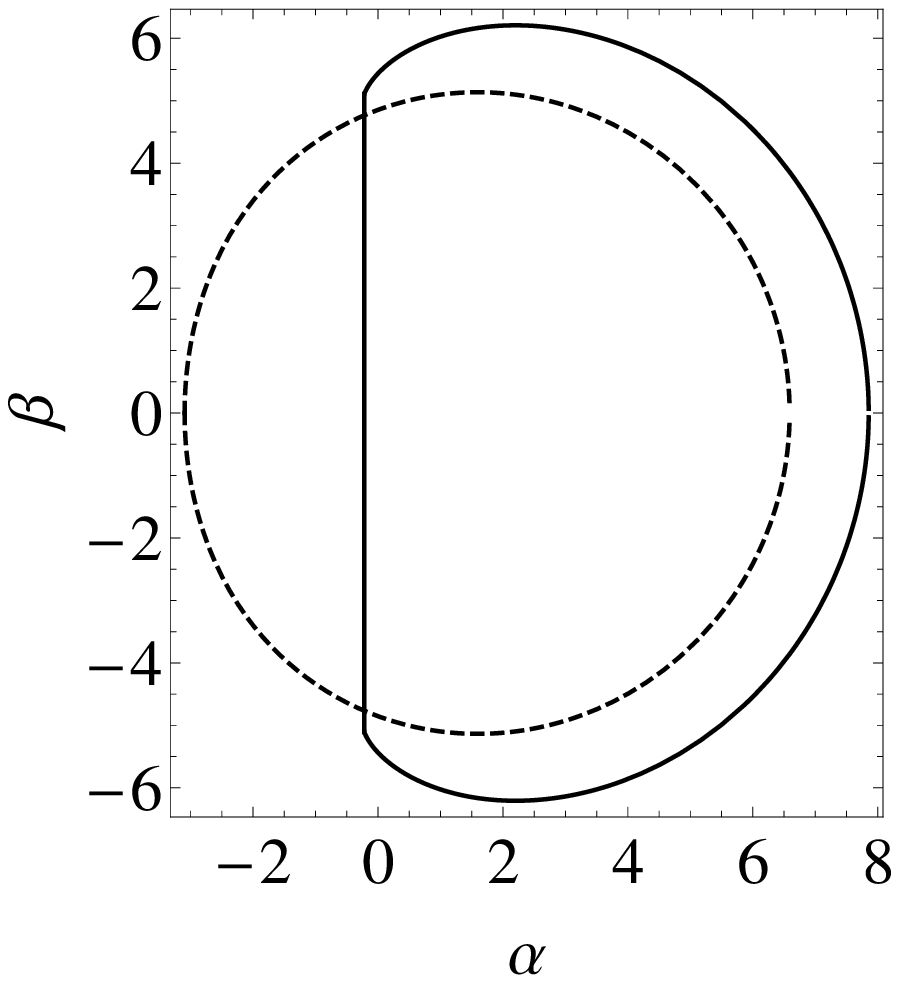} \\
			$J/M^{2}=0.01$, $\theta_{0}=60^\circ$\  &
			$J/M^{2}=0.05$, $\theta_{0}=60^\circ$\  &
            $J/M^{2}=0.3$, $\theta_{0}=60^\circ$\  &
			$J/M^{2}=0.9$, $\theta_{0}=60^\circ$ \\[2mm]
            \includegraphics[width=4.1cm]{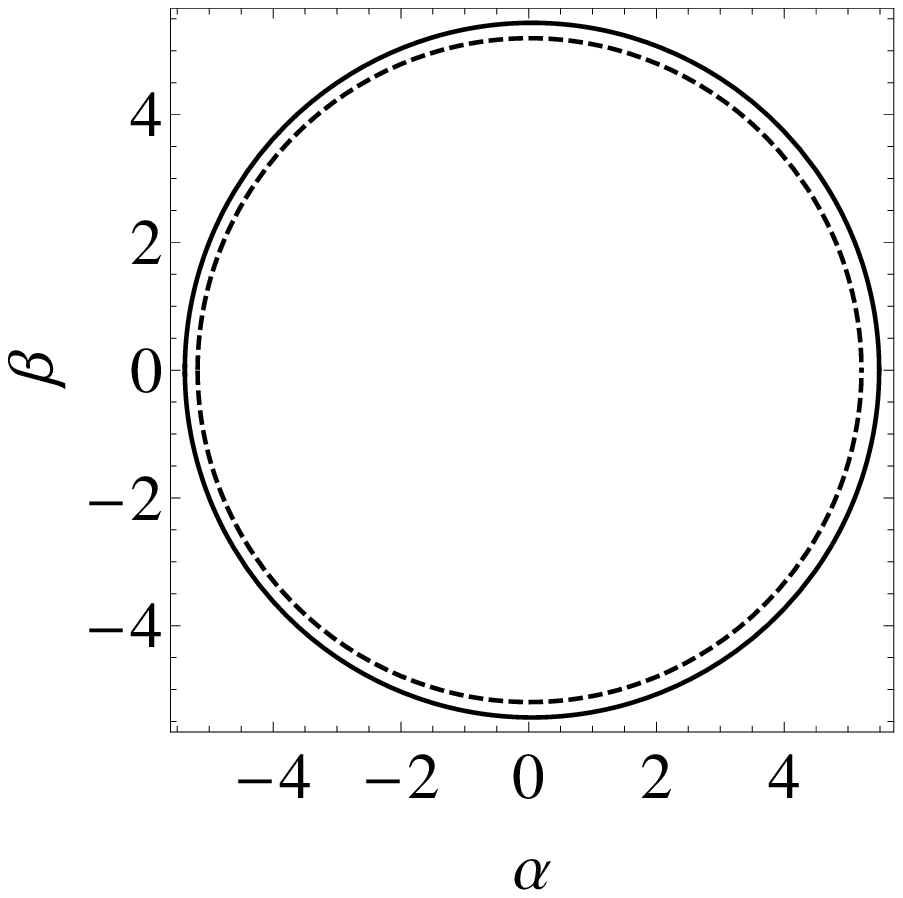} &
            \includegraphics[width=4.1cm]{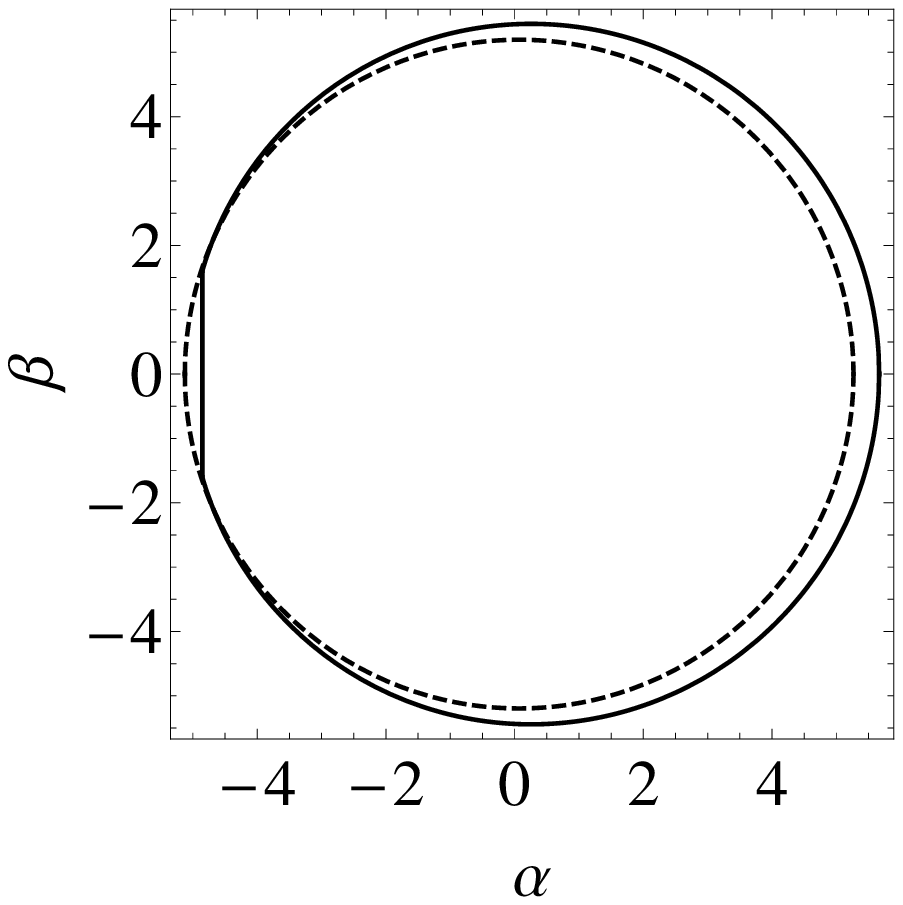} &
            \includegraphics[width=4.1cm]{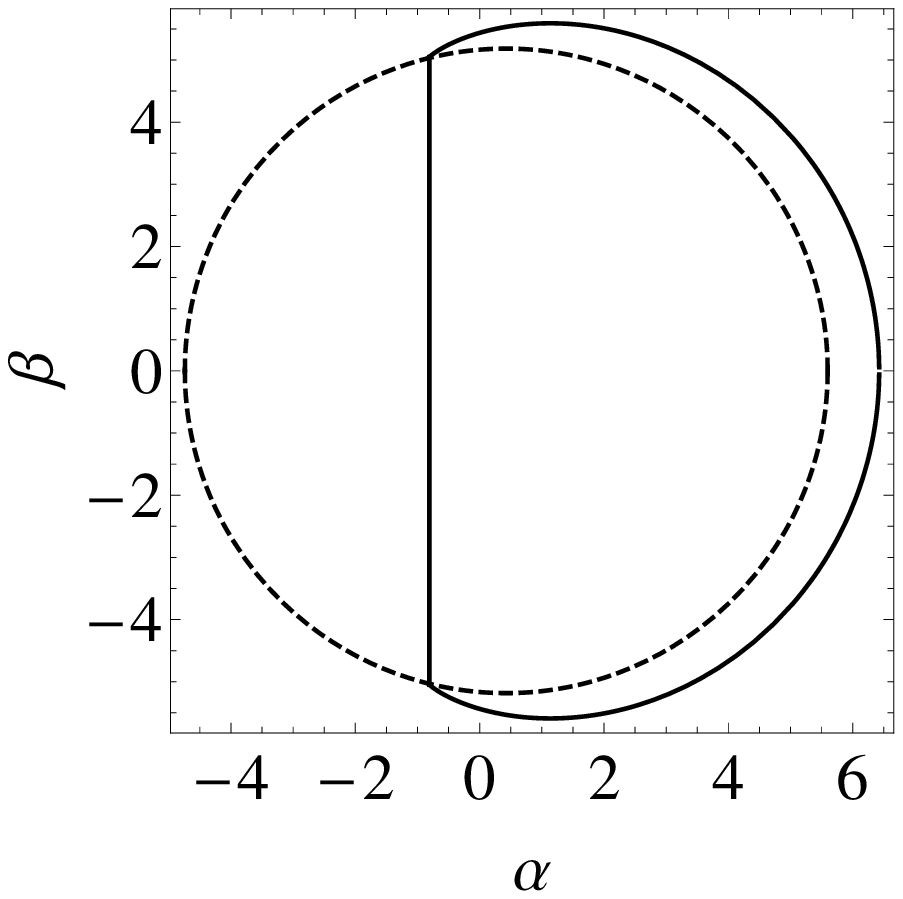} &
			\includegraphics[width=4.1cm]{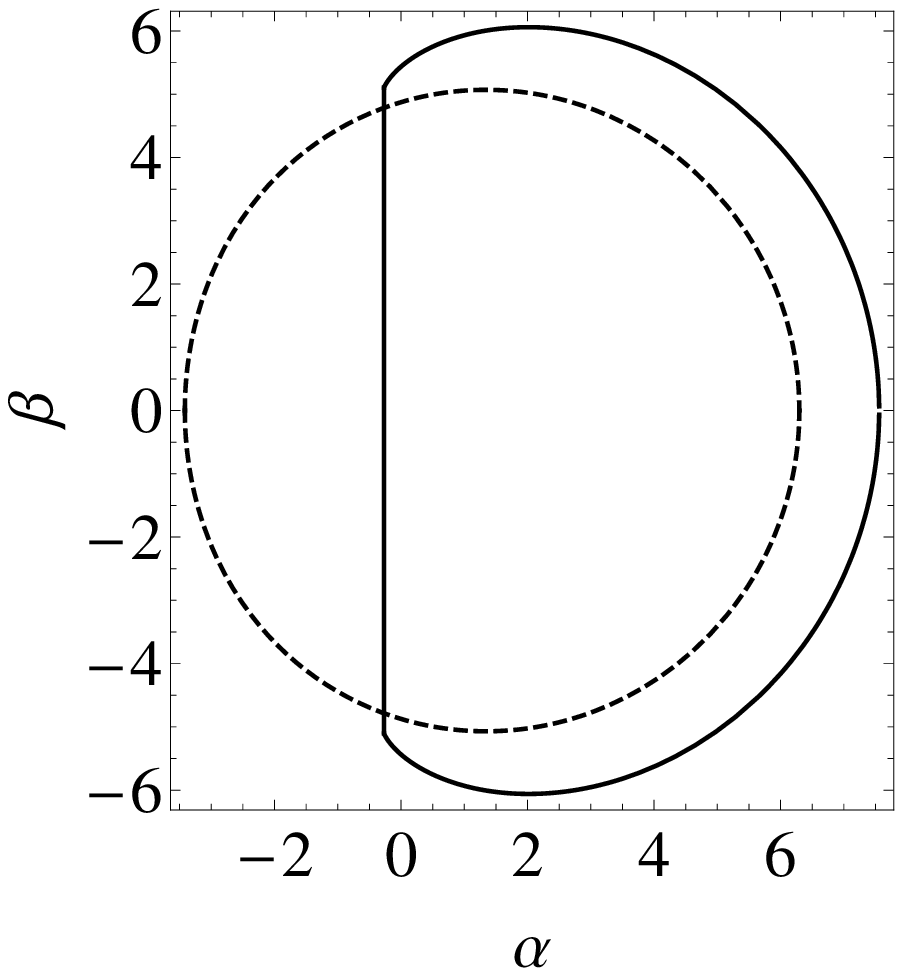} \\
			$J/M^{2}=0.01$, $\theta_{0}=45^\circ$\  &
			$J/M^{2}=0.05$, $\theta_{0}=45^\circ$\  &
            $J/M^{2}=0.3$, $\theta_{0}=45^\circ$\  &
			$J/M^{2}=0.9$, $\theta_{0}=45^\circ$ \\[2mm]
			\includegraphics[width=4.1cm]{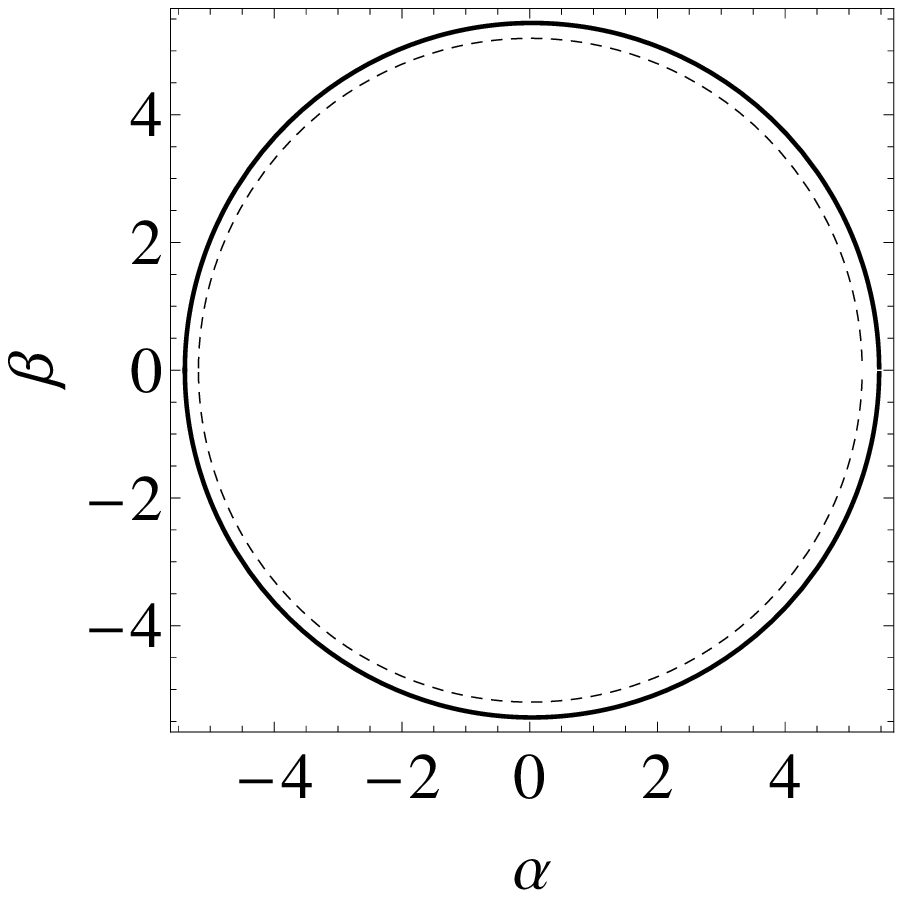} &
            \includegraphics[width=4.1cm]{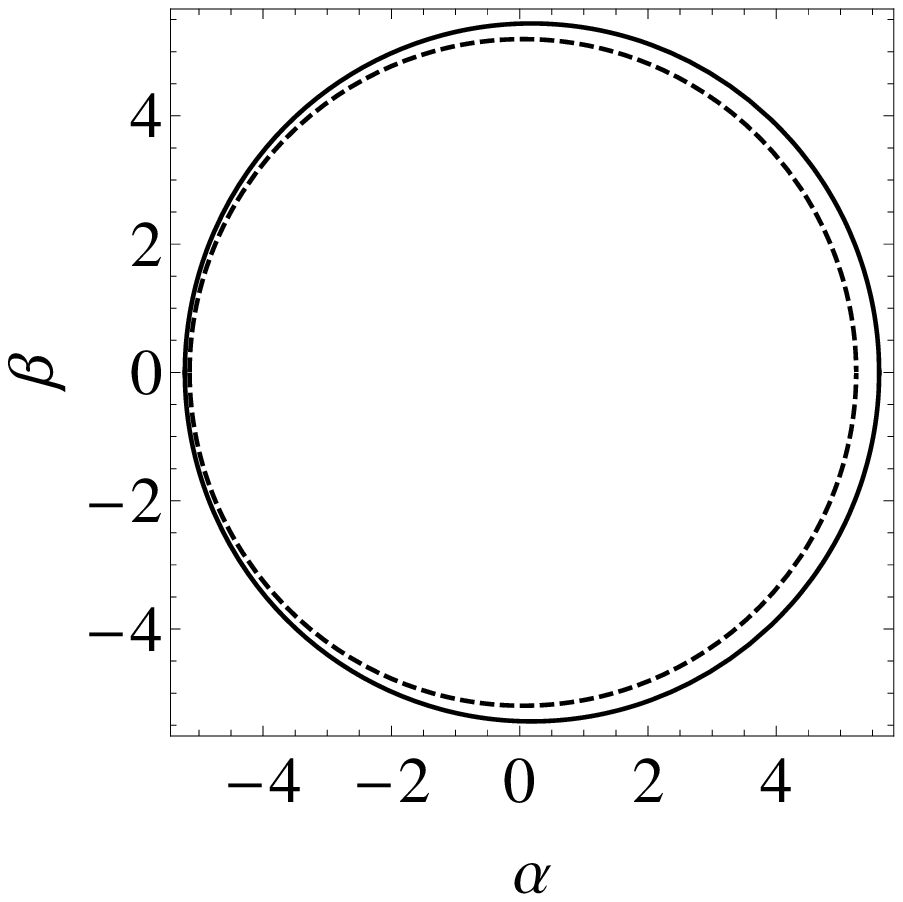} &
            \includegraphics[width=4.1cm]{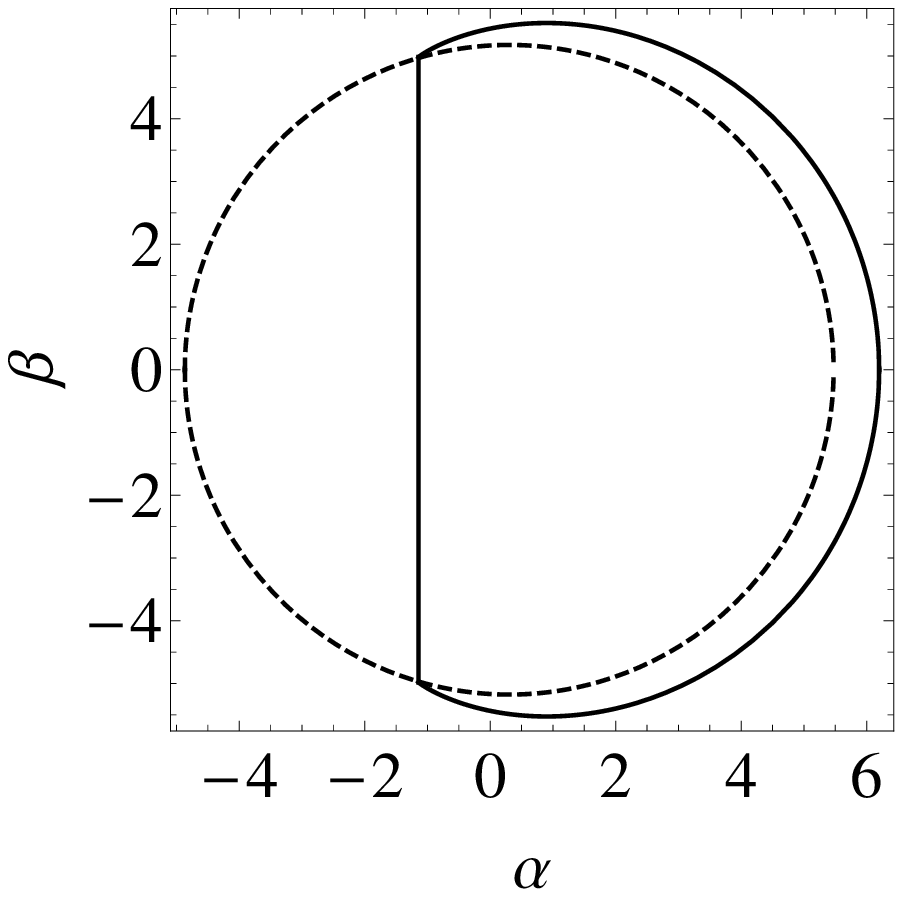} &
			\includegraphics[width=4.1cm]{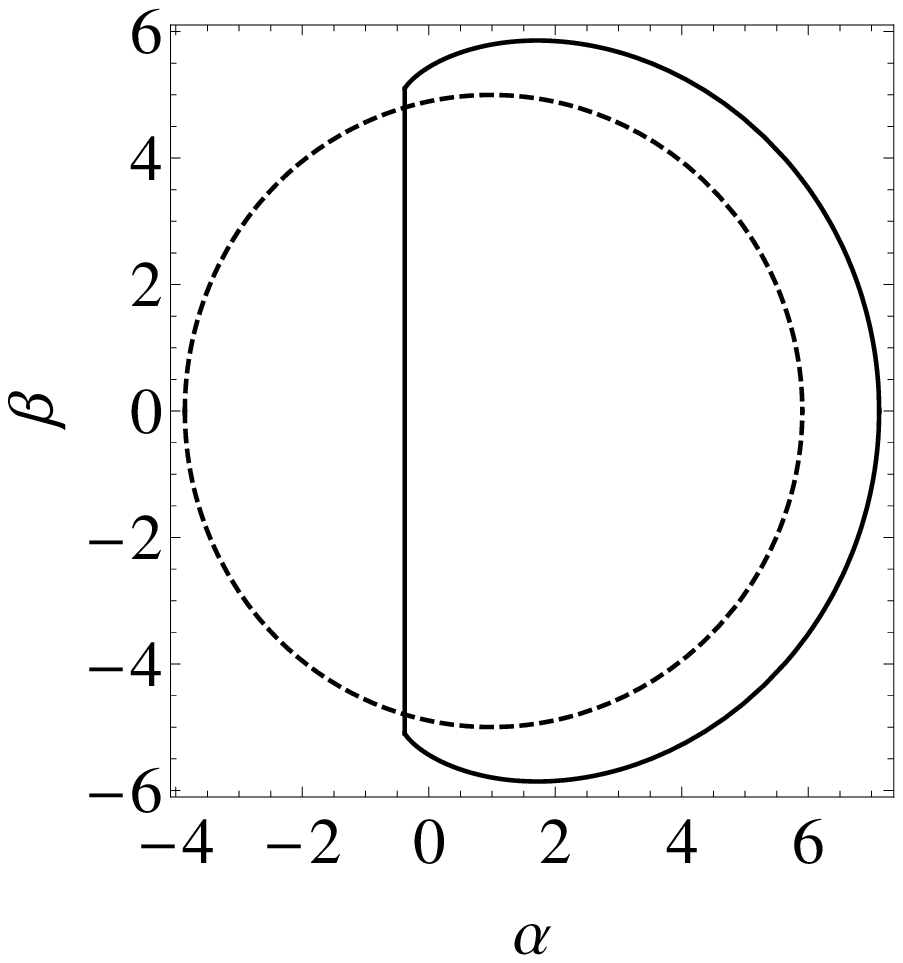} \\
			$J/M^{2}=0.01$, $\theta_{0}=30^\circ$\  &
			$J/M^{2}=0.05$, $\theta_{0}=30^\circ$\  &
            $J/M^{2}=0.3$, $\theta_{0}=30^\circ$\  &
			$J/M^{2}=0.9$, $\theta_{0}=30^\circ$ \\
		\end{tabular}}
\caption{\footnotesize{The shadow of rotating wormhole (solid line) and the Kerr black hole (dashed line) for different rotation parameters and inclination angles. The mass of both solutions is set equal to 1. The celestial coordinates ($\alpha, \beta$) are measured in the units of mass.}}
		\label{WS_a0}
\end{figure}

In the previous discussion we obtained analytical expressions for the boundary of the shadow for a general rotating wormhole described by ($\ref{metric}$). In order to be able to investigate the images, we should  consider a particular wormhole solution. The boundary of the shadow doesn't depend on the shape function $b(r)$, therefore we assume the simplest choice setting it equal to a constant $b=r_0>0$ which corresponds to the throat of the wormhole. The rest of the metric functions we choose in the form

\begin{eqnarray}\label{wormhole0}
N = \exp{\left(-\frac{r_0}{r}\right)}, \quad~~~ K =1,\quad~~~ \omega = \frac{2J}{r^3}.
\end{eqnarray}
The solution is parameterized by two parameters: $r_0$ which is equal to the mass of the wormhole $M$, and $J$ which is equal to its angular momentum. The shadow of this wormhole solution is presented in fig. \ref{WS_a0} for several inclination angles and angular momenta. We have set the mass of the wormhole $M =r_0 = 1$. For each set of parameters we have plotted also the shadow of the Kerr black hole with a dashed line for comparison. For small angular momenta the shadow of the wormhole is very similar to the Kerr black hole. However, increasing the angular momentum the shadow gets larger and the characteristic deformations of the image due to rotation becomes more distinctly expressed. In particular, the shift of the shadow to the right and the flattening of its left side  is stronger than in the Kerr case, and the two images start to deviate considerably.

\begin{figure}[htp]
		\setlength{\tabcolsep}{ 0 pt }{\scriptsize\tt
		\begin{tabular}{ cccc }
            \includegraphics[width=4.1cm]{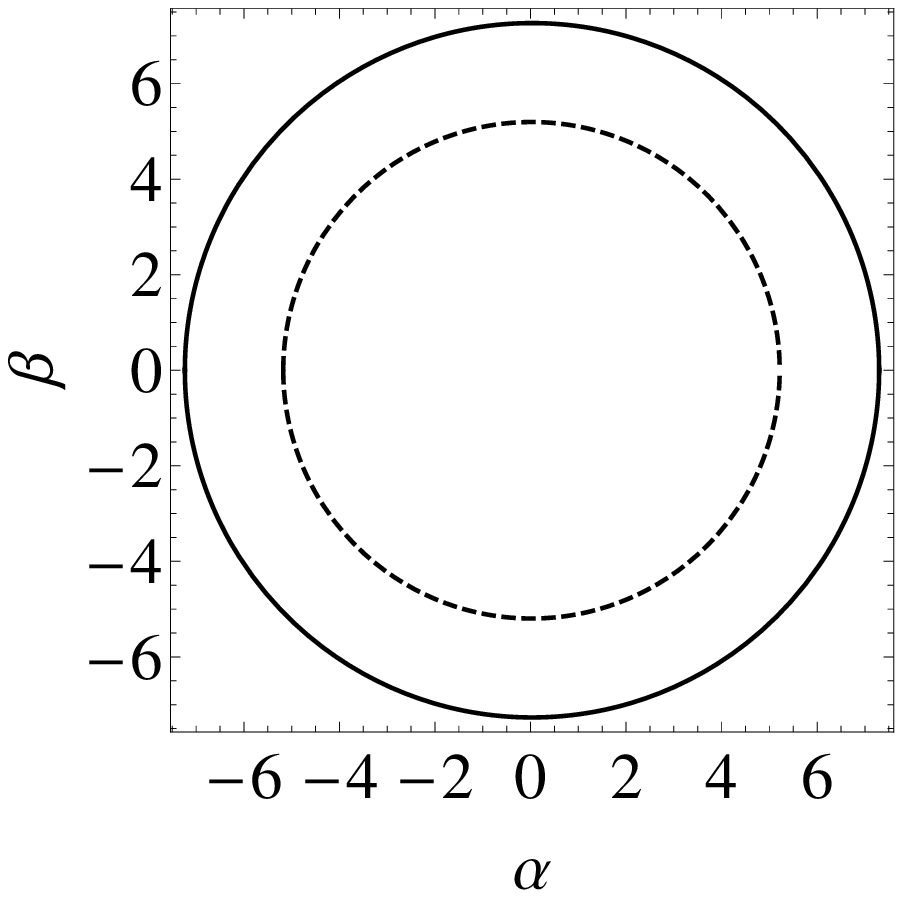} &
            \includegraphics[width=4.1cm]{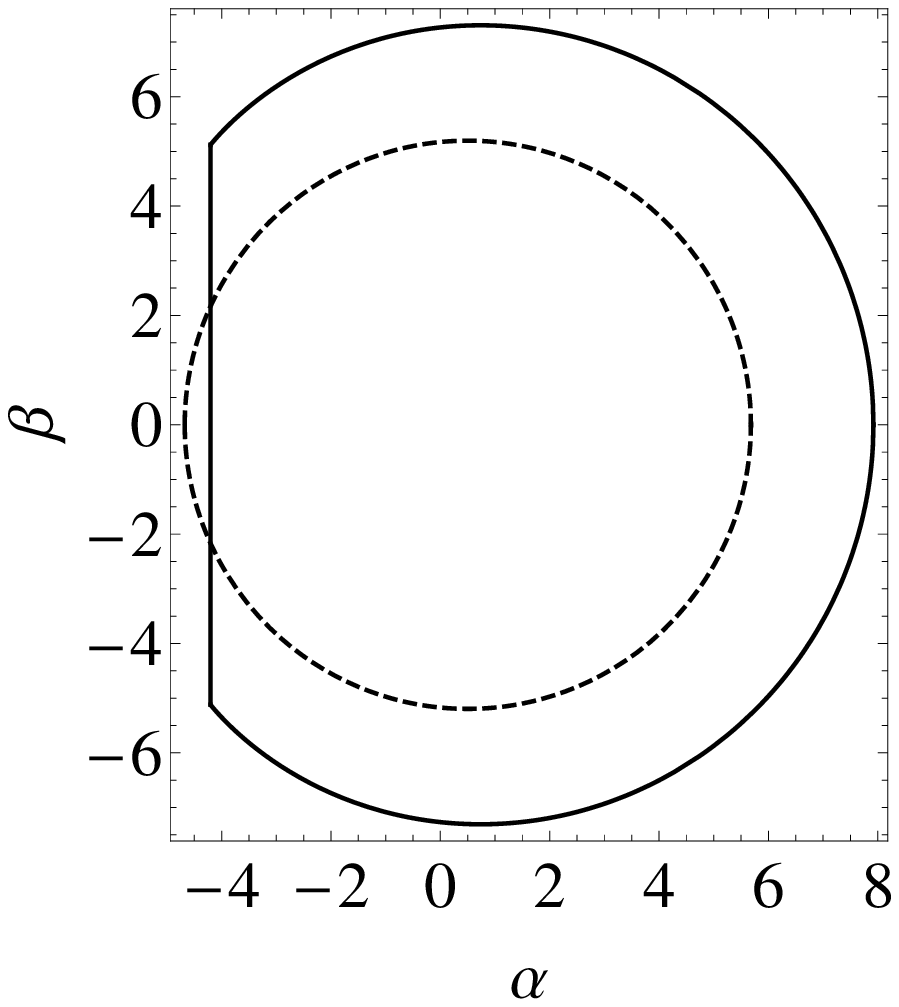} &
            \includegraphics[width=4.1cm]{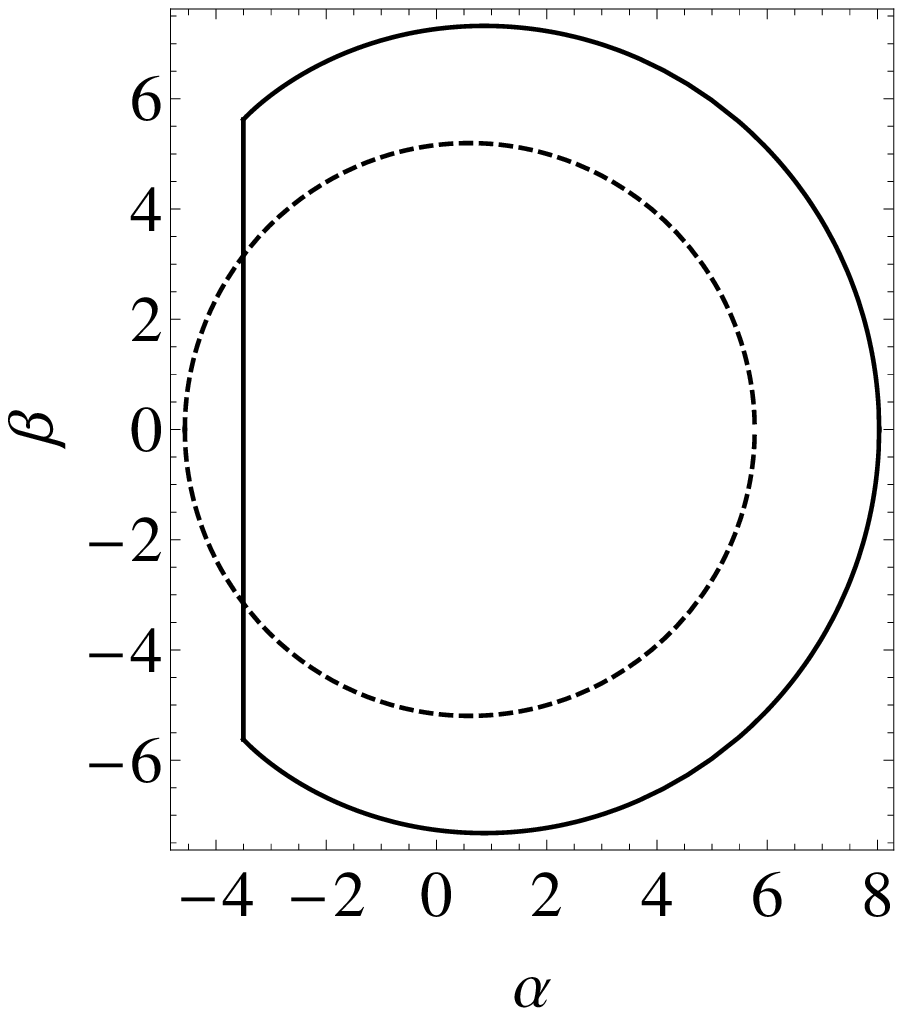} &
			\includegraphics[width=4.1cm]{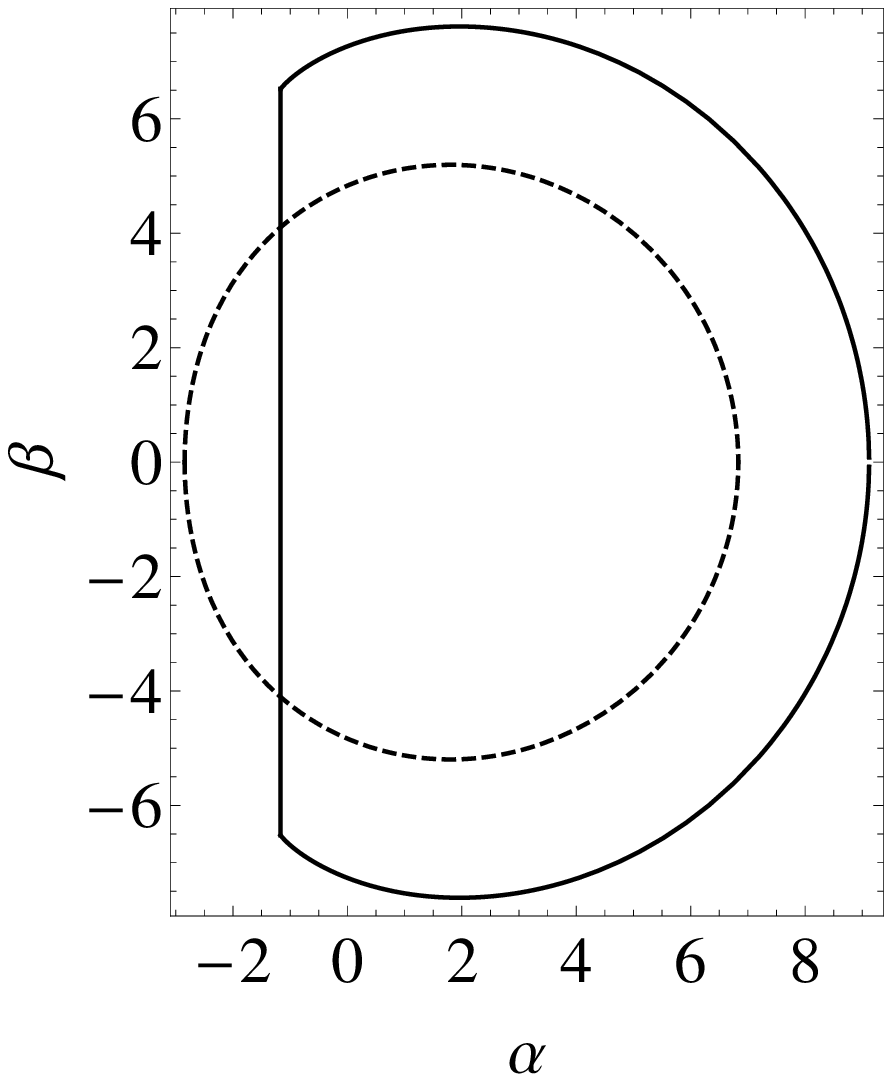} \\
			$J/M^{2}=0.01$, $\theta_{0}=90^\circ$;\  &
			$J/M^{2}=0.25$, $\theta_{0}=90^\circ$;\  &
            $J/M^{2}=0.3$, $\theta_{0}=90^\circ$;\  &
			$J/M^{2}=0.9$, $\theta_{0}=90^\circ$ \\
            		\end{tabular}}
\caption{\footnotesize The shadow of rotating wormhole with redshift function given by ($\ref{wormhole1}$) (solid line) and the Kerr black hole (dashed line) for different rotation parameters and inclination angles. The mass of both solutions is set equal to 1. The celestial coordinates ($\alpha, \beta$) are measured in the units of mass.}
		\label{WS_a1}
\end{figure}

We also consider another wormhole solution in which the redshift function is modified to
\begin{eqnarray}\label{wormhole1}
N = \exp{\left(-\frac{r_0}{r} - \frac{r_0^2}{r^2}\right)},
\end{eqnarray}
and the rest of the metric functions coincide with the previous case ($\ref{wormhole0}$). Its shadow is presented in fig. \ref{WS_a1} for
inclination angle $\theta_0 = \pi/2$ and several angular momenta, again compared to the shadow of the Kerr black hole. The two images can be
distinguished even in the static case ($J/M^2=0$), as the wormhole shadow is larger. By including rotation the distinction between the black hole
and the wormhole shadows gets more pronounced than in the case of the previous wormhole solution we considered. The same effects are observed for
other inclination angles as well.

\section{Conclusion}

The appearance of a shadow is a phenomenon which is not restricted only to black hole spacetimes. Under some circumstances it can be observed also by other compact objects as wormholes. We investigated a class of rotating traversable wormholes and obtained analytically the boundary of the shadow which they will cast. The images resemble the apparent shape of the Kerr black hole for small angular momenta, and get qualitatively distinct for large angular momenta.

\section*{Acknowledgements}
The support by the Bulgarian National Science Fund under Grant  DMU-03/6 is gratefully  acknowledged.


\begin{thebibliography}{tbds}

\bibitem{Flamm:1916}
L. Flamm,
Phys. Z., 17, 48 (1916).

\bibitem{Einstein:1935}
A. Einstein and N. Rosen,
 Phys. Rev. 48, 73 (1935).

\bibitem{Wheeler:1955}
J.A. Wheeler,
Phys. Rev. 97, 511 (1955).

\bibitem{Morris:1988}
M.~Morris, and K.~Thorne,
Am.\ J.\ Phys. {\bf56} (1988) 395.

\bibitem{Hawking:1973}
S.~Hawking, G.~Elis,
\textit{``The Large Scale Structure of the Spacetime"}, (Cambridge Univ. Press, 1973)


\bibitem{Teo:1998}
E.~Teo,
Phys.\ Rev.\ {\bf D58} (1998) 024014.

\bibitem{Kuhfittig:2003a}
P.~Kuhfittig
Phys.\ Rev.\ {\bf D67} (2003) 064015.

\bibitem{Visser:1989}
M. Visser,
Phys. Rev. D 39, 3182 (1989);
P.~Kuhfittig,
Phys.\ Rev.\ {\bf  D68} (2003) 067502;
A.~DeBenedictis and A.~Das,
Class. Quantum Grav. 18, 1187 (2001).




\bibitem{Sushkov:1992}
S. V. Sushkov,
Phys. Lett. A164, 33 (1992);
D. Hochberg, A. Popov and S. Sushkov,
Phys. Rev. Lett. 78, 2050 (1997);
R. Garattini,
Class. Quant. Grav. 22 1105 (2005).


\bibitem{Lobo:2005}
F.~Lobo,
 Phys. Rev. D71, 084011 (2005);
S. Sushkov,
 Phys. Rev. D71, 043520 (2005);
O. B. Zaslavskii,
 Phys. Rev. D72, 061303 (2005);
P. Kuhfittig,
Class. Quant. Grav. 23, 5853 (2006).

\bibitem{Kanti:2011}
P.~Kanti, B.~Kleihaus, and J.~Kunz,
Phys.Rev.Lett. {\bf107}, 271101  (2011);
P.~Kanti, B.~Kleihaus, and J.~Kunz,
Phys. Rev. D {\bf85}, 044007  (2012);
N.~Montelongo~Garc\'{\i}a, and F.~Lobo,
Class. Quantum Grav. {\bf28}, 085018 (2011);
T.~ Harko, F.~ Lobo, M.~Mak, S.~ Sushkov
Phys.\ Rev.\ {\bf D87} (2013) 067504;

\bibitem{Cramer:1995}
J. G. Cramer, R. L. Forward, M. S. Morris, M. Visser,
G. Benford and G. A. Landis, Phys. Rev. D 51,
3117 (1995) ;  F. Abe, Astrophys. J. 725, 787
(2010);  N. Tsukamoto,T. Harada and K. Yajima,
Phys. Rev. D 86, 104062(2012)

\bibitem{Muller:2008}
T. M\"{u}ller,
Phys. Rev. D 77, 044043 (2008);
A.~Abdujabbarov, B.~Ahmedov,
Astrophys.\ Space\ Sci. {\bf 321} (2009) 225.
V. Kagramanova, E. Smolarek,
"Dynamics of test particles in thin-shell wormhole spacetimes",
[arXiv:1304.5646].

\bibitem{Harko:2008}
T. Harko, Z. Kovacs and F. Lobo,
Phys. Rev. D 78, 084005 (2008);
T. Harko, Z. Kovacs and F. Lobo,
Phys.\ Rev.\ {\bf D79} (2009) 064001.

\bibitem{Falke:2000}
H. Falcke, F. Melia, and E. Agol,
Astrophys.J. 528 (2000) L13;
J.-P. Luminet,
Astron. Astrophys. 75  (1979) 228.


\bibitem{Bardeen}
 J. Bardeen, \textit{Black Holes}, Edited by C. De Witt and B.S. De Witt, \'Ecole d' \'et\'e de Physique Th\'eorique, Les Houches 1972 (Gordon and Breach Science Publishers, New York, 1973).

\bibitem{Chandra}
S. Chandrasekhar, \textit{The mathematical theory of black holes} (Oxford Univ. Press, 1992).

\bibitem{Frolov:2011}
V.~Frolov, A.~Zelnikov, \textit{Inroduction to black hole physics} (Oxford Univ. Press, 2011)

\bibitem{Young:1976}
P.~J.~Young,
Phys.\ Rev.\  D {\bf 14}, 3281 (1976).

\bibitem{de Vries}
A. de Vries, Class. Quant. Grav. {\bf 17}, 123 (2000).


\bibitem{Takahashi}
R. Takahashi, Astrophys. J. {\bf 611} (2004) 996 .

\bibitem{Bambi}
C. Bambi, and K. Freese,
Phys. Rev. D {\bf79} (2009) 043002.

\bibitem{Hioki:2009}
K.~Hioki, K.~Maeda,
Phys.\ Rev.\ {\bf D80}, 024042 (2009)

\bibitem{Bambi:2013}
C. Bambi,
Phys. Rev. D {\bf87} (2013) 107501.

\bibitem{EH}
http://www.eventhorizontelescope.org.

\bibitem{RS}
http://www.asc.rssi.ru/radioastron/

\bibitem{Johannsen:2012}
T. Johannsen, D. Psaltis, S. Gillessen, D.P. Marrone, F. EOzel, S.S. Doeleman, and V.L. Fish,
Astrophys. J. 758, 30 (2012).

\bibitem{MX}
http://bhi.gsfc.nasa.gov/maxim-home.html


\bibitem{Ahmedov:2013}
A. Abdujabbarov, F. Atamurotov, Y. Kucukakca, B. Ahmedov, U. Camci,
Astrophys. Space Sci. 344 (2013) 429.

\bibitem{Amarilla:2013}
L.~Amarilla, E.~Eiroa
Phys.Rev. {\bf D87} (2013) 044057

\bibitem{Amarilla:2010}
L.~Amarilla, E.~Eiroa, G.~Giribet,
Phys.\ Rev.\ {\bf D81} (2010) 124045.

\bibitem{Hioki:2008}
K. Hioki and U. Miyamoto,
Phys. Rev. {\bf D78} (2008) 044007.

\bibitem{Carter:1968}
B.~Carter,
Phys.\ Rev.\  {\bf 174}, 1559 (1968).

\bibitem{Bray:1986}
I.~Bray,
Phys.\ Rev. {\bf D34} (1986) 367;
S. Vazquez and E. Esteban,
Nuovo Cim. {\bf 119B} (2004) 489.


\end{thebibliography}
\end{document}